\input amstex
\documentstyle{amsppt}
%
\catcode`@=11
\redefine\output@{%
  \def\break{\penalty-\@M}\let\par\endgraf
  \ifodd\pageno\global\hoffset=105pt\else\global\hoffset=8pt\fi  
  \shipout\vbox{%
    \ifplain@
      \let\makeheadline\relax \let\makefootline\relax
    \else
      \iffirstpage@ \global\firstpage@false
        \let\rightheadline\frheadline
        \let\leftheadline\flheadline
      \else
        \ifrunheads@ 
        \else \let\makeheadline\relax
        \fi
      \fi
    \fi
    \makeheadline \pagebody \makefootline}%
  \advancepageno \ifnum\outputpenalty>-\@MM\else\dosupereject\fi
}
\catcode`\@=\active
\nopagenumbers
\def\negskp{\hskip -2pt}
\def\divr{\operatorname{div}}
\def\rot{\operatorname{rot}}
\def\const{\operatorname{const}}
\def\grad{\operatorname{grad}}
\accentedsymbol\ty{\tilde y}
\def\blue#1{#1}
\font\tencyr=wncyr10
\pagewidth{360pt}
\pageheight{606pt}
\leftheadtext{Jeffrey Comer, Ruslan Sharipov}
\rightheadtext{A note on the kinematics of dislocations.}
\topmatter
\title
A note on the kinematics of dislocations in crystals.
\endtitle
\author
Jeffrey Comer, Ruslan Sharipov
\endauthor
\address Physics Department, The University of Akron, Akron, OH 
\endaddress
\email jeffcomer\@gmail.com
\endemail
\address Mathematics Department, Bashkir State University,
450074 Ufa, Russia
\endaddress
\email \vtop to 30pt{\hsize=280pt\noindent
R\_\hskip 1pt Sharipov\@ic.bashedu.ru\newline
r-sharipov\@mail.ru\newline
ra\_\hskip 1pt sharipov\@lycos.com\vss}
\endemail
\urladdr
http:/\negskp/www.geocities.com/r-sharipov
\endurladdr
\abstract
    A part of the theory of dislocations in crystals is revised with the
aim to fit it into the framework of the nonlinear theory of plasticity
initially designed for amorphous glassy materials.
\endabstract
\endtopmatter
\loadbold
\TagsOnRight
\document

\head
1. Geometry of dislocations.
\endhead
    Dislocations provide a microscopic mechanism explaining the plasticity
of crystals. The idea of dislocations suggested by Taylor and Orowan in 1934 
\vadjust{\vskip 5pt\hbox to 0pt{\kern -10pt
\includegraphics{disl01.eps}\hss}\vskip 320pt}is illustrated on
the first two figures. Fig\.~1.1 and Fig\.~1.2 show screw-type and
edge-type dislocations respectively. There are also mixed type dislocations
combining the features of both screw and edge dislocations.\par
    In the continuous limit, a single dislocation can be understood as shown on Fig\.~1.3 and Fig\.~1.4 above. Imagine a smooth curve $AB$ within the continuous
medium being the edge of some surface $S$. This fact is denoted as $AB=
\partial S$. Imagine that the medium is cut along the surface $S$ and then glued with some displacement (see Fig\.~1.4). Upon gluing, all points of the
surface $S$ outside the dislocation line $AB$ become regular points of the medium (as regular as all other points within the medium). Though a
dislocation produces the stress and elastic deformation around itself, it
doesn't produce the defects of crystalline grid outside the dislocation line $AB$. Therefore, if we cut out a sufficiently small spherical neighborhood
$G$ of some point $C\notin AB$ and then release it, we get some stress-free
crystalline body with no defects of the crystalline grid (see Fig\.~1.5 and Fig\.~1.6).
\vadjust{\vskip 5pt\hbox to 0pt{\kern -10pt
\includegraphics{disl02.eps}\hss}\vskip 160pt}Mathematically 
this fact is expressed by a map from the spherical neighborhood $G$ of 
the point $C$ to some domain $\Omega$ in $\Bbb R^3$ (see Fig\.~1.6). This
map can be given by three functions 
$$
\hskip -2em
\cases x^1=x^1(y^1,y^2,y^3),\\
x^2=x^2(y^1,y^2,y^3),\\
x^3=x^3(y^1,y^2,y^3).
\endcases
\tag1.1
$$
Here $y^1,\,y^2,\,y^3$ are some curvilinear coordinates within the real 
crystalline body, while $x^1,\,x^2,\,x^3$ are Cartesian coordinates 
in the three-dimensional space $\Bbb R^3$ which we used in our mental 
experiment where we cut out the ball $G$. The Jacobi matrix
$$
\hskip -2em
\hat T^{\,i}_j=\frac{\partial x^i}{\partial y^j}
\tag1.2
$$
of the map \thetag{1.1} is non-degenerate since otherwise it would mean
the infinite compressibility of the medium. The map $f\!:\,G\to\Omega$
is invertible and due to the non-degeneracy condition $\det\hat\bold T
\neq 0$, the inverse map $f^{-1}\!:\,\Omega\to G$ is also given by three
smooth functions similar to \thetag{1.1}:
$$
\hskip -2em
\cases y^1=y^1(x^1,x^2,x^3),\\
y^2=y^2(x^1,x^2,x^3),\\
y^3=y^3(x^1,x^2,x^3).
\endcases
\tag1.3
$$
By differentiating \thetag{1.3} we find the inverse Jacobi matrix 
$\hat\bold S=\hat\bold T^{-1}$:
$$
\hskip -2em
\hat S^{\,i}_j=\frac{\partial y^i}{\partial x^j}.
\tag1.4
$$\par
    Note that the map \thetag{1.1} can be extended to any connected
and simply connected domain within the crystalline body that comprises 
the point $C$ and does not comprise the points of dislocation, line
$AB$. Such an extension is non-degenerate ($\det\hat\bold T\neq 0$), 
but in the general case it is not globally bijective. In such a case, 
the inverse map \thetag{1.3} is defined only locally. However, the components of the mutually inverse Jacobi matrices \thetag{1.2} and \thetag{1.4} can be treated as global functions
$$
\xalignat 2
&\hskip -2em
\hat T^{\,i}_j=\hat T^{\,i}_j(y^1,y^2,y^3),
&&\hat S^{\,i}_j=\hat S^{\,i}_j(y^1,y^2,y^3)
\tag1.5
\endxalignat
$$
defined at all points of the crystal except for those lying on the
\vadjust{\vskip 5pt\hbox to 0pt{\kern -10pt
\includegraphics{disl03.eps}\hss}\vskip 155pt}dislocation 
line $AB$.\par
    Now let's consider a crystal with one dislocation line $AB$ and
define the following path integral along some closed path $\gamma$ 
that encircles the dislocation line $AB$:
$$
\hskip -2em
b^{\,i}=\oint\limits_\gamma\sum^3_{j=1}\hat T^{\,i}_j\,\tau^j\,ds
\tag1.6
$$
(see Fig\.~1.7). Here $\tau^1,\,\tau^2\,\tau^3$ are the components of 
the tangent vector $\boldsymbol\tau$ of the path $\gamma$. For the path
given parametrically by three functions 
$$
\xalignat 3
&y^1=y^1(s),&&y^2=y^2(s),&&y^3=y^3(s)
\endxalignat
$$
this tangent vector is defined by three derivatives
$$
\tau^j=\frac{dy^j}{ds},\quad j=1,\,2,\,3.
$$\par
    Note that the path integral \thetag{1.6} along any path that does not
encircle the dislocation line (see Fig\.~1.7) is identically zero:
$$
\hskip -2em
\oint\limits_\mu\sum^3_{j=1}\hat T^{\,i}_j\,\tau^j\,ds=0.
\tag1.7
$$
Indeed, due to \thetag{1.2} the integral \thetag{1.7} is transformed
into the path integral of the second kind applied to the total differential
of the smooth function $x^i(y^1,y^2,y^3)$:
$$
\oint\limits_\mu\sum^3_{j=1}\hat T^{\,i}_j\,\tau^j\,ds=
\oint\limits_\mu dx^i(y^1,y^2,y^3)=0.
$$\par
\proclaim{Theorem 1.1} The value of the integral \thetag{1.6} is an 
invariant of the dislocation $AB$. It does not depend on a particular 
contour $\gamma$ encircling the dislocation line.
\endproclaim
    The proof of this theorem is obvious from Fig\.~1.8. Indeed, for the
pair of contours $\gamma_1$ and $\gamma_2$ on Fig\.~1.8 we have
$$
\oint\limits_{\gamma_1}\sum^3_{j=1}\hat T^i_j\,\tau^j\,ds-
\oint\limits_{\gamma_2}\sum^3_{j=1}\hat T^i_j\,\tau^j\,ds=
\oint\limits_\mu\sum^3_{j=1}\hat T^i_j\,\tau^j\,ds=0.
$$
\definition{Definition 1.1} Three constants $b^1,\,b^2,\,\,b^3$ determined
by the integral \thetag{1.6} are the components of a vector $\bold b$
characterizing the dislocation line. This vector is called the {\it Burgers
vector} of a dislocation.
\enddefinition    
    Note that the Burgers vector $\bold b$ is not a vector in the space of
real crystalline body. It is associated with the imaginary space of 
stress-free crystalline matter shown on Fig\.~1.6. In what follows we shall
call this space the {\it Burgers space}.\par
    The concept of Burgers space is convenient for understanding the nature
of matrices \thetag{1.5}. Although they have upper and lower indices and
depend on the coordinates of a point in the real crystalline body, they are
not components of traditional tensor fields. They are {\it double space
tensors}. The index $j$ in $\hat T^i_j$ is a covariant index associated with
the space of real crystalline body, while $i$ is a contravariant index
associated with the Burgers space. As for the inverse matrix $\hat\bold S=
\hat\bold T^{-1}$ in \thetag{1.5}, its lower index $j$ is associated with the
Burgers space, while its upper index is related to the real crystalline
body.\par
     Usually each dislocation line is a closed path within crystalline body.
Otherwise, if it is not closed, it should begin at some point on the boundary
of the crystalline body and it should end at some other point which is on
the boundary \vadjust{\vskip 5pt\hbox to 0pt{\kern -10pt
\includegraphics{disl04.eps}\hss}\vskip 155pt}as well. Usually,
each dislocation line is taken with some orientation assigned to it. One
can change the orientation of a dislocation line, however, in this case its
Burgers vector $\bold b$ is changed for the opposite one: $\bold b\to 
-\bold b$. If the closed contour $\gamma$ encircles several dislocation
lines (see Fig\.~1.9), then
$$
\hskip -2em
\oint\limits_\gamma\sum^3_{j=1}\hat T^i_j\,\tau^j\,ds=
\pm b^{\,i}(1)\pm\ldots\pm b^{\,i}(N).
\tag1.8
$$
The sign of each Burgers vector in right hand side of \thetag{1.8} is
determined by the orientation of corresponding dislocation line.\par
    In some cases dislocation lines have brunching points as shown on
Fig\.~1.10. For the Burgers vectors of dislocation lines in this case
we have the equality
$$
\hskip -2em
\bold b_1=\bold b_2+\bold b_3.
\tag1.9
$$
The equality \thetag{1.9} is an analog of Kirchhoff rule for currents
in electromagnetism. Its proof is clear from Fig\.~1.10.
\head
2. Continual limit for dislocations.
\endhead
     In order to detect macroscopic phenomena associated with dislocations
we should have a substantial amount of dislocations in each macroscopically
essential volume of the medium. In this case, instead of considering individual dislocation lines, we consider the density of Burgers vectors
for dislocations
$$
\hskip -2em
\rho^{\,i}_j=\rho^{\,i}_j(y^1,y^2,y^3).
\tag2.1
$$
Like $\bold S$ and $\bold T$ in \thetag{1.5}, the functions \thetag{2.1} 
are components of a double space tensorial field. The upper index $i$ in
\thetag{2.1} is associated with the Burgers space, while $j$ is a
traditional tensorial index associated with the space of real crystalline
body.\par
     Since dislocation lines cannot end in the interior of a crystal and
since they obey the conservation law \thetag{1.9} at their  brunching
points, the amount of Burgers vectors flowing into some domain trough
its boundary with dislocation lines is equal to the amount of this vector
flowing out of this domain. This fact is written as the following integral
equality for the density of Burgers vectors \thetag{2.1}:
$$
\hskip -2em
\int\limits_{\partial\Omega}\sum^3_{j=1}\rho^{\,i}_j\,n^j\,dS=0.
\tag2.2
$$
Here $n^1,\,n^3,\,n^3$ are components of the unit vector $\bold n$ of
the external normal to the boundary $\partial\Omega$ of the domain $\Omega$.
The differential form of the equality \thetag{2.2} looks like
$$
\hskip -2em
\sum^3_{j=1}\sum^3_{k=1}g^{kj}\,\nabla_{\!k}\rho^{\,i}_j=0.
\tag2.3
$$
The covariant derivative $\nabla_{\!k}\rho^{\,i}_j$ in the formula
\thetag{2.3} is calculated as follows:
$$
\hskip -2em
\nabla_{\!k}\rho^{\,i}_j=\frac{\partial\rho^{\,i}_j}{\partial y^k}
-\sum^3_{q=1}\Gamma^q_{kj}\,\rho^{\,i}_q.
\tag2.4
$$
Note that $g^{kj}$ in \thetag{2.3} are components of the metric tensor
and $\Gamma^q_{kj}$ in \thetag{2.4} are components of the metric connection.
They are determined by the choice of curvilinear coordinates
$y^1,\,y^2,\,y^3$ (see 
\catcode`#=11\catcode`#=6\blue{\cite{1}}%
). Note also that in writing \thetag{2.4}, we
do not apply the standard rule of covariant differentiation to the upper
index $i$. This is because it is associated with the Burgers space other
than the space of real crystalline body. In short form the equality
\thetag{2.3} is written as 
$$
\hskip -2em
\divr\boldsymbol\rho=0.
\tag2.5
$$
In this form \thetag{2.5} the equality \thetag{2.3} resembles the Maxwell
equation $\divr\bold H=0$ in electromagnetism. However, one should remember
the difference: $\bold H$ is a vector, while $\boldsymbol\rho$ in \thetag{2.5} is a double space tensor.\par
\parshape 14 0pt 360pt 0pt 360pt 180pt 180pt 180pt 180pt 180pt 180pt 
180pt 180pt 180pt 180pt 180pt 180pt 180pt 180pt 180pt 180pt 180pt 180pt 
180pt 180pt 180pt 180pt 0pt 360pt
     Let $S$ be some surface spanned onto the contour $\gamma$. Then
$\gamma=\partial S$ (see Fig\.~2.1). The total flow of Burgers vectors
\vadjust{\vskip 5pt\hbox to 0pt{\kern -10pt
\includegraphics{disl05.eps}\hss}\vskip -5pt}across the 
surface $S$ is given by the following surface integral (compare with \thetag{2.2}):
$$
\hskip -2em
b^i=\int\limits_S\sum^3_{j=1}\rho^{\,i}_j\,n^j\,dS.
\tag2.6
$$
The value of the integral \thetag{2.6} cannot\linebreak change unless 
some dislocation lines move and cross the contour $\gamma$. For this reason 
the time derivative $db^i/dt$ is given by some path integral along
the contour $\gamma$:
$$
\hskip -2em
\frac{db^i}{dt}=-\oint\limits_\gamma\sum^3_{k=1} j^{\,i}_k\,\tau^k\,ds.
\tag2.7
$$
The double space tensorial quantity $j^{\,i}_k$ in \thetag{2.7} 
is called the {\it current of Burgers vectors} produced by moving 
dislocations. Combining \thetag{2.6} and \thetag{2.7}, we get
$$
\hskip -2em
\int\limits_S\sum^3_{j=1}\frac{\partial\rho^{\,i}_j}{\partial t}
\,n^j\,dS+\oint\limits_{\partial S}\sum^3_{k=1} j^{\,i}_k\,\tau^k\,ds=0.
\tag2.8
$$
The integral equation \thetag{2.8} can be transformed to differential form by applying the Stokes formula. As a result we obtain the following
equation:
$$
\hskip -2em
\frac{\partial\rho^{\,i}_k}{\partial t}+\sum^3_{q=1}\sum^3_{p=1}
\sum^3_{r=1}\sum^3_{m=1}\omega_{kqp}\ g^{qr}\,g^{pm}\,\nabla_{\!r}
j^{\,i}_m=0.
\tag2.9
$$
Here $g^{qr}$ and $g^{pm}$ are the components of metric tensor, and
$\omega_{kqp}$ are the components of a completely skew-symmetric tensor.
It is called the {\it volume tensor}. Its components are expressed
through Levi-Civita symbol:
$$
\omega_{kqp}=\sqrt{\det(g_{ij})}\,\varepsilon_{kqp}
$$
(see more details in 
\catcode`#=11\catcode`#=6\blue{\cite{1}}%
). When applying the covariant derivative 
$\nabla_{\!r}$ to a double space tensor one should remember that 
the indices associated with the Burgers space are ignored. For
the derivative $\nabla_{\!r}j^{\,i}_m$ in \thetag{2.9} we have
$$
\hskip -2em
\nabla_{\!r}j^{\,i}_m=\frac{\partial j^{\,i}_m}{\partial y^r}-
\sum^3_{s=1}\Gamma^s_{rm}j^{\,i}_s
\tag2.10
$$
(compare \thetag{2.10} and \thetag{2.4}). Like the equation \thetag{2.5}
above, the differential equation \thetag{2.9} can be written in a shorter form:
$$
\hskip -2em
\frac{\partial\boldsymbol\rho}{\partial t}+
\rot\bold j=0.
\tag2.11
$$
Like the equation \thetag{2.5}, this equation \thetag{2.11} is an analog
of corresponding Max\-well equation in electromagnetism (see 
\catcode`#=11\catcode`#=6\blue{\cite{2}}%
):
$$
\frac{1}{c}\,\frac{\partial\bold H}{\partial t}+\rot\bold E=0.
$$\par
     Now let's return to the matrices \thetag{1.5}. For a single
dislocation they are determined by formulas \thetag{1.2} and \thetag{1.4}. Usually they are singular functions at the points of dislocation line, just
like Coulomb potential of a point charge. However, if charges are treated as
continuously smeared in the space, the electric potential
$\varphi(t,y^1,y^2,y^3)$ is a smooth function. Similarly, in continuous 
limit of dislocation theory $\hat T^i_j(t,y^1,y^2,y^3)$ and 
$\hat S^i_j(t, y^1,y^2,y^3)$ are smooth functions forming two matrices inverse to each other. In this case they are not determined by formulas
\thetag{1.2} and \thetag{1.4} any more. Instead, we have the equality
$$
\hskip -2em
\oint\limits_{\partial S}\sum^3_{j=1}\hat T^i_j\,\tau^j\,ds=
\int\limits_S\sum^3_{j=1}\rho^{\,i}_j\,n^j\,dS
\tag2.12
$$
derived from \thetag{1.8}. Applying the Stokes formula to \thetag{2.12}, 
we get
$$
\hskip -2em
\rot\hat\bold T=\boldsymbol\rho.
\tag2.13
$$
In coordinate form the equation \thetag{2.13} is written as follows:
$$
\sum^3_{q=1}\sum^3_{p=1}
\sum^3_{r=1}\sum^3_{m=1}\omega_{kqp}\ g^{qr}\,g^{pm}\,\nabla_{\!r}
\hat T^{\,i}_m=\rho^{\,i}_k.
$$
Comparing \thetag{2.13} with $\bold H=\rot\bold A$, we conclude that the 
double space tensor field $\hat\bold T$ with components 
$\hat T^i_j(t,y^1,y^2,y^3)$ here plays the same role as the vector-potential
$\bold A$ in electromagnetism.\par
\head
3. Deformation tensors.
\endhead
    Let's consider the motion of a crystalline medium in the presence
of dislocations in it. Here we reproduce in part the content of 
\catcode`#=11\catcode`#=6\blue{\cite{3}}%
 in order to have the same notations. Suppose that a
point of the medium with coordinates $\ty^1,\,\ty^2,\,\ty^3$ has moved to
the point with coordinates $y^1,\,y^2,\,y^3$. Then we have a map:
$$
\hskip -2em
\cases y^1=y^1(t,\ty^1,\ty^2,\ty^3),\\
y^2=y^2(t,\ty^1,\ty^2,\ty^3),\\
y^3=y^3(t,\ty^1,\ty^2,\ty^3).
\endcases
\tag3.1
$$
This is the displacement map $\tau$. The argument $t$ in \thetag{3.1} 
is responsible for the time evolution of the displacement. The time 
derivatives of the functions \thetag{3.1} determine the components of
the velocity vector $\bold v$:
$$
\hskip -2em
v^i=\dot y^i=\frac{\partial y^i}{\partial t},\qquad
i=1,2,3.
\tag3.2
$$
As defined in \thetag{3.2}, $v^1,\,v^2\,\,v^3$ are the functions of
$t,\,\ty^1,\,\ty^2,\,\ty^3$. However, in order to interpret them as the
components of a vector field, they should depend on the coordinates 
of the current actual position of a point of the medium. To change
the arguments of the derivatives \thetag{3.2} we use the inverse
displacement map $\tau^{-1}$:
$$
\hskip -2em
\cases \ty^1=\ty^1(t,y^1,y^2,y^3),\\
\ty^2=\ty^2(t,y^1,y^2,y^3),\\
\ty^3=\ty^3(t,y^1,y^2,y^3).
\endcases
\tag3.3
$$
The time dependent maps \thetag{3.1} and \thetag{3.4} define two Jacobi
matrices $\tilde S$ and $\tilde T$:
$$
\xalignat 2
&\hskip -2em
\tilde S^{\,i}_j=\frac{\partial y^i}{\partial\ty^j},
&&\tilde T^{\,i}_j=\frac{\partial\ty^i}{\partial y^j}.
\tag3.4
\endxalignat
$$
The nonlinear deformation tensor $\bold G$ then is defined as follows
(see 
\catcode`#=11\catcode`#=6\blue{\cite{3}}%
):
$$
\hskip -2em
G_{ij}=\sum^3_{r=1}\sum^3_{s=1}g_{rs}(\ty^1,\ty^2,\ty^3)\ \tilde T^r_i
\,\tilde T^s_j.
\tag3.5
$$
Upon transforming all arguments in \thetag{3.5} into $t,\,y^1,\,y^2,
\,y^3$ we get a tensor field $\bold G$ with components $G_{ij}=G_{ij}
(t,\,y^1,\,y^2,\,y^3)$. Differentiating \thetag{3.5}, by direct 
calculations one can derive the following formula:
$$
\hskip -2em
\frac{\partial G_{ij}}{\partial t}+\sum^3_{k=1}v^k\,\nabla_{\!k}G_{ij}=
-\sum^3_{k=1}G_{kj}\,\nabla_{\!i}v^k-\sum^3_{k=1}G_{ik}\,\nabla_{\!j}v^k.
\tag3.6
$$
In order to describe the plasticity of amorphous materials in 
\catcode`#=11\catcode`#=6\blue{\cite{3}}%
 the following decomposition of the deformation tensor 
$\bold G$ was suggested:
$$
\hskip -2em
G_{ij}=\sum^3_{k=1}\sum^3_{q=1}\check G^{\,k}_i\,\hat G_{kq}
\,\check G^{\,q}_j.
\tag3.7
$$
Here $\hat G_{kq}$ are components of the elastic deformation tensor
$\hat\bold G$, while $\check G^{\,k}_i$ and $\check G^{\,q}_j$
are components of the plastic deformation tensor $\check\bold G$.
For these tensor fields $\hat\bold G$ and $\check\bold G$ in 
\catcode`#=11\catcode`#=6\blue{\cite{3}}%
 the following evolution equations were suggested:
$$
\gather
\hskip -2em
\gathered
\frac{\partial\hat G_{kq}}{\partial t}+\sum^3_{r=1}v^r\,
\nabla_{\!r}\hat G_{kq}=-\sum^3_{r=1}\nabla_{\!k}v^r\,
\hat G_{rq}-\sum^3_{r=1}\hat G_{kr}\,\nabla_{\!q}v^r+\\
+\sum^3_{r=1}\theta^{\,r}_k\,\hat G_{rq}+\sum^3_{r=1}
\hat G_{kr}\,\theta^{\,r}_q.
\endgathered
\tag3.8\\
\vspace{2ex}
\hskip -2em
\frac{\partial\check G^{\,k}_i}{\partial t}+
\sum^3_{r=1}v^r\,\nabla_{\!r}\check G^{\,k}_i=
\sum^3_{r=1}\left(\check G^{\,r}_i\,\nabla_{\!r}v^k
-\nabla_{\!i}v^r\,\check G^{\,k}_r\right)-\sum^3_{r=1}
\theta^{\,k}_r\,\check G^{\,r}_i.
\tag3.9
\endgather
$$
The main goal of the present paper is to show that the decomposition 
\thetag{3.7} and the differential equations \thetag{3.8} and
\thetag{3.9} can be consistently incorporated into the existing
theory of plasticity in crystals in its nonlinear version.\par
\head
4. Kinematics of a dislocated medium.
\endhead
    Let's begin with the equation \thetag{2.11} and substitute 
\thetag{2.13} into it. As a result we obtain the following differential
equation:
$$
\rot\left(\frac{\partial\hat\bold T}{\partial t}+\bold j\,\right)=0.
$$
It is known that a vectorial field with zero curl is the gradient of some
scalar field:
$$
\hskip -2em
\frac{\partial\hat\bold T}{\partial t}+\bold j=-\grad\bold w.
\tag4.1
$$
In our case all of the fields $\hat\bold T$, $\bold j$ and $\bold w$ in
\thetag{4.1} are double space tensors; they have one upper index 
associated with the Burgers space. Therefore, \thetag{4.1} is written
as
$$
\hskip -2em
\frac{\partial\hat T^i_k}{\partial t}+j^{\,i}_k=-\frac{\partial w^i}
{\partial y^k}.
\tag4.2
$$
The vector $\bold w$ in \thetag{4.2} can be interpreted as the velocity
vector (it easy to check that its components $w^1,\,w^2,\,w^3$ are
measured in $cm\cdot sec^{-1}$). However, $\bold w\neq\bold v$. Indeed,
the components of the velocity vector $\bold v$ defined in \thetag{3.2}
and then transformed into the arguments $t,\,y^1,\,y^2,\,y^3$ by means of
the map \thetag{3.3} represent a traditional tensor field with one upper
index, while $\bold w$ is a double space tensor field. Below we shall
understand $\bold w$ as an independent parameter of a dislocated medium.
The physical nature of this parameter is not yet clear to us, it will
be studied in separate paper. However, there is a transparent analogy 
with electromagnetism:
$$
\frac{1}{c}\,\frac{\partial\bold A}{\partial t}+
\bold E=-\grad\varphi.
$$
Comparing \thetag{4.1} with this equality, we find that $\bold w$ is
an analog of the scalar potential $\varphi$ of electromagnetic field.
Moreover, this equality supports our previous associations of $\hat
\bold T$ with $\bold A$, $\bold j$ with $\bold E$ and $\boldsymbol
\rho$ with $\bold H$.\par
     Suppose that the initial state of our crystalline medium is free 
of dislocations. Below we assume that it is stress-free too. Then
$$
\xalignat 2
&\hskip -2em
\rho^{\,i}_k(0,\ty^1,\ty^2,\ty^3)=0,
&&j^{\,i}_k(0,\ty^1,\ty^2,\ty^3)=0.
\tag4.3
\endxalignat
$$
Due to \thetag{4.3} we can arrange the bijective map from the space of the
real crystalline body to the Burgers space. It is given by three functions
$$
\hskip -2em
\cases x^1=x^1(\ty^1,\ty^2,\ty^3),\\
x^2=x^2(\ty^1,\ty^2,\ty^3),\\
x^3=x^3(\ty^1,\ty^2,\ty^3).
\endcases
\tag4.4
$$
The map \thetag{4.4} is an isometry because we assume that the initial 
state of the crystal has no deformation. The isometry condition is
written as
$$
\hskip -2em
g_{rs}(\ty^1,\ty^2,\ty^3)=
\sum^3_{p=1}\sum^3_{q=1}\overset\sssize\star\to g\vphantom{g}_{pq}
\ \overset\sssize\star\to T\vphantom{T}^p_r\ \overset\sssize\star
\to T\vphantom{T}^q_s.
\tag4.5
$$
Here $\overset\sssize\star\to g\vphantom{g}_{pq}$ are the components of
metric tensor in the Burgers space, \vadjust{\vskip -3pt} while $\overset
\sssize\star\to T\vphantom{T}^p_r$ and $\overset\sssize\star\to
T\vphantom{T}^q_s$ are the components of Jacobi matrix for the map
\thetag{4.4}:
$$
\overset\sssize\star\to T\vphantom{T}^p_r=\frac{\partial x^p}
{\partial\ty^r}.
$$
Note that $\overset\sssize\star\to g\vphantom{g}_{pq}=\const$ since we
choose Cartesian coordinates in the Burgers space (see Fig\.~1.6).\par
     The next step is to add the time variable to the map \thetag{4.4}.
For this purpose let's use the inverse evolution map \thetag{3.3} and
let's consider the composite map
$$
\hskip -2em
\cases x^1=x^1(\ty^1(t,y^1,y^2,y^3),\,\ldots,\,\ty^3(t,y^1,y^2,y^3)),\\
x^2=x^2(\ty^1(t,y^1,y^2,y^3),\,\ldots,\,\ty^3(t,y^1,y^2,y^3)),\\
x^3=x^3(\ty^1(t,y^1,y^2,y^3),\,\ldots,\,\ty^3(t,y^1,y^2,y^3)).
\endcases
\tag4.6
$$
Using the chain rule, for the Jacobi matrix of the map \thetag{4.6} we
write
$$
\hskip -2em
T^p_r=\frac{\partial x^p}{\partial y^i}=\sum^3_{r=1}
\overset\sssize\star\to T\vphantom{T}^p_r\ 
\tilde T^r_i.
\tag4.7
$$
From \thetag{3.5}, \thetag{4.5}, and \thetag{4.7} one easily derives
$$
\hskip -2em
G_{ij}=\sum^3_{r=1}\sum^3_{s=1}\overset\sssize\star\to g\vphantom{g}_{pq}
\ T^p_i\ T^q_j.
\tag4.8
$$
The equality \thetag{4.8} means that the deformation tensor $\bold G$
can be defined through the composite map \thetag{4.6}. As for the matrix
\thetag{4.7}, we interpret $T^p_r$ as the components of a double space 
tensor $\bold T$. Both tensors $\bold T$ and $\hat\bold T$ are called the
{\it distorsion tensors} (see 
\catcode`#=11\catcode`#=6\blue{\cite{4--7}}%
): $\bold T$ represents the
{\it compatible distorsion} since it is given by partial derivatives in
\thetag{4.7}, while $\hat\bold T$ represents the {\it incompatible
distorsion} since the equality \thetag{1.2} is not valid upon passing 
to the continuous limit (see \thetag{2.12} and \thetag{2.13}, see also the comment above the formula \thetag{2.12}).\par
     The compatible distorsion arises due to the macroscopic deformation 
of a crystal. In the elastic case, the macroscopic deformation is transferred
to the microscopic level and produces the same distorsion of interatomic
bonds (see Fig\.~4.1 and Fig\.~4.2). \vadjust{\vskip 5pt\hbox to 0pt{\kern
-10pt\includegraphics{disl06.eps}\hss}\vskip 110pt}The plastic
deformation is that very case, when some interatomic bonds get torn and
then relinked in a different way. On Fig\.~4.3 we see the birth of a pair
of the edge dislocations with mutually opposite Burgers vectors. On 
Fig\.~4.4, Fig\.~4.5, and Fig\.~4.6 one of them moves from the left to the
right. \vadjust{\vskip 5pt\hbox to 0pt{\kern -10pt\includegraphics{disl07.eps}\hss}\vskip 110pt}Behind the moving dislocation the series of undistorted cells arises, while the total (macroscopic) distortion angle remains unchanged. This fact explains why $\hat\bold T\neq\bold T$ for plastic deformations.\par
     The elastic response of a body is determined by the elongation and/or
contraction of interatomic bonds within it. For this reason let's define the
elastic deformation tensor $\hat\bold G$ by analogy with \thetag{4.8}, but
using $\hat\bold T$ instead of $\bold T$:
$$
\hskip -2em
\hat G_{ij}=\sum^3_{p=1}\sum^3_{q=1}\overset\sssize\star\to
g\vphantom{g}_{pq}\ \hat T^p_i\ \hat T^q_j.
\tag4.9
$$
The components of the plastic deformation tensor $\check\bold G$ are defined 
as follows:
$$
\hskip -2em
\check G^{\,k}_i=\sum^3_{p=1}\hat S^k_p\ T^p_i.
\tag4.10
$$
Here $\hat S^k_p$ are components of the inverse matrix $\hat\bold S=\hat
\bold T^{-1}$. Though defined through the double space tensors, the
deformation tensors $\hat\bold G$ and $\check\bold G$ are traditional 
tensor fields in the space of the real crystalline body. The indices $p$
and $q$ associated with the Burgers space in \thetag{4.9} and 
\thetag{4.10} both are summation indices. They disappear when the
sums are evaluated.\par
     Note that now the decomposition \thetag{3.7} follows immediately
from the expressions \thetag{4.8}, \thetag{4.9}, and \thetag{4.10}.
Therefore, it is sufficient to derive the equation \thetag{3.9}. The
equation \thetag{3.8} then is derived from \thetag{3.9} and \thetag{3.6}
due to the decomposition \thetag{3.7}. As the first step in deriving
the differential equation \thetag{3.9}, we differentiate the equality
\thetag{4.10} with respect to the time variable $t$:
$$
\hskip -2em
\frac{\partial\check G^{\,k}_i}{\partial t}=\sum^3_{p=1}
\frac{\partial\hat S^k_p}{\partial t}\ T^p_i+\sum^3_{p=1}
\hat S^k_p\ \frac{\partial T^p_i}{\partial t}.
\tag4.11
$$
In order to differentiate the inverse matrix $\hat\bold S=\hat\bold T^{-1}$
in formula \thetag{4.11}, we use the well-known standard formula $\hat\bold
S'=-\hat\bold S\ \hat\bold T'\ \hat\bold S$:
$$
\hskip -2em
\frac{\partial\hat S^k_p}{\partial t}=-
\sum^3_{q=1}\sum^3_{r=1}
\hat S^k_q\ \frac{\partial\hat T^q_r}{\partial t}\ \hat S^r_p.
\tag4.12
$$
Substituting \thetag{4.12} into \thetag{4.11} and applying the
formula \thetag{4.2}, now we derive
$$
\hskip -2em
\frac{\partial\check G^{\,k}_i}{\partial t}=\sum^3_{q=1}
\sum^3_{r=1}\hat S^k_q\left(j^q_r+\frac{\partial w^q}
{\partial y^r}\right)\check G^r_i+\sum^3_{p=1}
\hat S^k_p\ \frac{\partial T^p_i}{\partial t}.
\tag4.13
$$
For the time derivative of $T^p_i$ in formula \thetag{4.13} we have 
the following equality:
$$
\hskip -2em
\frac{\partial T^p_i}{\partial t}=-\sum^3_{r=1}
\frac{\partial\left(v^r\,T^p_r\right)}{\partial y^i}
\tag4.14
$$
The equality \thetag{4.14} is derived in few steps by applying 
the chain rule to the mapping functions \thetag{4.6} and to their
inverse mapping functions
$$
\cases y^1=y^1(t,\ty^1(x^1,x^2,x^3),\,\ldots,\,\ty^3(x^1,x^2,x^3)),\\
y^2=y^2(t,\ty^1(x^1,x^2,x^3),\,\ldots,\,\ty^3(x^1,x^2,x^3)),\\
y^3=y^3(t,\ty^1(x^1,x^2,x^3),\,\ldots,\,\ty^3(x^1,x^2,x^3)).
\endcases
$$
Apart from \thetag{4.14} we need the identity
$$
\hskip -2em
\frac{\partial T^p_r}{\partial y^i}=
\frac{\partial T^p_i}{\partial y^r},
\tag4.15
$$
which follows immediately from \thetag{4.7}. Now, applying the formulas
\thetag{4.14} and \thetag{4.15} to \thetag{4.13}, we derive the following
equality:
$$
\hskip -2em
\gathered
\frac{\partial\check G^{\,k}_i}{\partial t}=\sum^3_{q=1}
\sum^3_{r=1}\hat S^k_q\left(j^{\,q}_r+\frac{\partial w^q}
{\partial y^r}\right)\check G^{\,r}_i-\sum^3_{r=1}
\frac{\partial v^r}{\partial y^i}\,\check G^{\,k}_r\,-\\
-\sum^3_{p=1}\sum^3_{r=1}\hat S^k_p\ v^r\,
\frac{\partial T^p_i}{\partial y^r}.
\endgathered
\tag4.16
$$
The last term in \thetag{4.16} can be transformed as follows:
$$
\gather
\sum^3_{p=1}\sum^3_{r=1}\hat S^k_p\ v^r\,\frac{\partial T^p_i}{\partial y^r}
=\sum^3_{p=1}\sum^3_{r=1}v^r\,\frac{\partial\bigl(\hat S^k_p\
T^p_i\,\bigr)}{\partial y^r}-\sum^3_{p=1}\sum^3_{r=1}v^r\,
\frac{\partial\hat S^k_p}{\partial y^r}\ T^p_i=\\
=\sum^3_{r=1}v^r\,\frac{\partial\check G^{\,k}_i}{\partial y^r}+
\sum^3_{q=1}\sum^3_{p=1}\sum^3_{r=1}S^k_q\,v^r\,\frac{\partial\hat T^q_p}
{\partial y^r}\ \check G^{\,p}_i.
\endgather
$$
Upon substituting this expression into \thetag{4.16} we derive
$$
\hskip -2em
\gathered
\frac{\partial\check G^{\,k}_i}{\partial t}+
\sum^3_{r=1}v^r\,\frac{\partial\check G^{\,k}_i}{\partial y^r}=
\sum^3_{r=1}\left(\check G^{\,r}_i\,\frac{\partial v^k}{\partial y^r}
-\frac{\partial v^r}{\partial y^i}\,\check G^{\,k}_r\right)-\\
-\sum^3_{r=1}\left(\frac{\partial v^k}{\partial y^r}+
\sum^3_{q=1}\hat S^k_q\left(-j^{\,q}_r-\frac{\partial w^q}
{\partial y^r}+\sum^3_{p=1}v^p\,\frac{\partial\hat T^q_r}
{\partial y^p}\right)\!\!\right)\check G^{\,r}_i.
\endgathered
\tag4.17
$$
Now let's introduce the following notations:
$$
\hskip -2em
\theta^{\,k}_r=
\frac{\partial v^k}{\partial y^r}+
\sum^3_{q=1}\hat S^k_q\left(-j^{\,q}_r-\frac{\partial w^q}
{\partial y^r}+\sum^3_{p=1}v^p\,\frac{\partial\hat T^q_r}
{\partial y^p}\right).
\tag4.18
$$
Then the differential equation \thetag{4.17} for plastic deformation
tensor is rewritten as
$$
\hskip -2em
\frac{\partial\check G^{\,k}_i}{\partial t}+
\sum^3_{r=1}v^r\,\frac{\partial\check G^{\,k}_i}{\partial y^r}=
\sum^3_{r=1}\left(\check G^{\,r}_i\,\frac{\partial v^k}{\partial y^r}
-\frac{\partial v^r}{\partial y^i}\,\check G^{\,k}_r\right)-
\sum^3_{r=1}\theta^{\,k}_r\,\check G^{\,r}_i.
\tag4.19
$$
By direct calculations one can verify that all of the partial derivatives
in \thetag{4.19} can be replaced by covariant derivatives. As a result
\thetag{4.19} takes the form of the equation \thetag{3.9}, which was
derived in 
\catcode`#=11\catcode`#=6\blue{\cite{3}}%
 for amorphous materials.\par
     The partial derivatives in formula \thetag{4.18} can also be replaced 
by covariant ones. As a result this formula is rewritten as
$$
\hskip -2em
\theta^{\,k}_r=\nabla_{\!r}v^k-\sum^3_{q=1}\hat S^k_q\,j^{\,q}_r
-\sum^3_{q=1}\hat S^k_q\,\nabla_{\!r}w^q
+\sum^3_{q=1}\sum^3_{p=1}v^p\,\hat S^k_q\,\nabla_{\!p}\hat T^q_r.
\tag4.20
$$
Indeed, if we remember that
$$
\xalignat 3
&\nabla_{\!r}v^k=\frac{\partial v^k}{\partial y^r}+\sum^3_{p=1}
\Gamma^k_{rp}\,v^p,
&&\nabla_{\!r}w^q=\frac{\partial w^q}{\partial y^r},
&&\nabla_{\!p}\hat T^q_r=\frac{\partial\hat T^q_r}{\partial y^p}
-\sum^3_{m=1}\Gamma^m_{rp}\,\hat T^q_m
\endxalignat
$$
(see \thetag{2.4} and \thetag{2.10} for comparison), we see that the
Christoffel symbols $\Gamma^k_{rp}$ do cancel each other when substituting
the above expressions into \thetag{4.20}.\par
    The formula \thetag{4.20} written in terms of covariant derivatives reveals the tensorial nature of the quantities $\theta^{\,k}_r$: they are
the components of a traditional tensor field $\boldsymbol\theta$ (not a
double space tensor unlike $\hat\bold T$, $\boldsymbol\rho$, and 
$\bold j$).\pagebreak
\proclaim{Theorem 4.1} The nonlinear deformation tensor $\bold G$ in
the theory of crystalline dislocations admits the decomposition \thetag{3.7}
into elastic and plastic parts $\hat\bold G$ and $\check\bold G$, both
satisfying the same differential equations \thetag{3.8} and \thetag{3.9}
as in the theory of plasticity for amorphous materials.
\endproclaim
\head
5. Conclusions.
\endhead
     The theorem~4.1 proved by the above calculations is the main result
of present paper. It is a purely kinematic (i\.\,e\. geometric) result.
Among other results, one should mention the concept of the Burgers space.
The interpretation of Burgers vectors as the vectors of a separate space
(and, hence, the use of double space tensors) is our methodical achievement
(in linear theory this feature is completely hidden, see 
\catcode`#=11\catcode`#=6\blue{\cite{8}}%
). The further development of this technique and the
further comparison of amorphous and crystalline plasticity theories
(including the dynamics and thermodynamics of media) will be done in separate papers.
\head
6. Acknowledgments from Jeffrey Comer.
\endhead
I would like to thank S. F. Lyuksyutov and The University of Akron's Department of Physics for allowing a short course on tensors within his electromagnetism course.
\head
7. Acknowledgments from Ruslan Sharipov.
\endhead
I am grateful to Yu\.~A.~Osipyan whose lectures on dislocation 
physics I attended in the early 1980s along with other MIPT\footnotemark\ students of the group \#\,728. I am also grateful to V.~Ya\.~Kravchenko 
who was my scientific adviser at that time.\linebreak I am especially grateful to
S.~F.~Lyuksyutov who got me back to the research in condensed matter 
physics after more than 20 years in pure mathematics.\par
\footnotetext{\ Moscow Institute of Physics and Technology, see
\blue{http:/\negskp/www.mipt.ru} in Russian
or\linebreak%
\blue{http:/\negskp/www.phystech.edu} in English.}

\enddocument
\end